\def\year{2020}\relax
\documentclass[letterpaper]{article} % DO NOT CHANGE THIS
\usepackage{aaai20}  % DO NOT CHANGE THIS
\usepackage{times}  % DO NOT CHANGE THIS
\usepackage{helvet} % DO NOT CHANGE THIS
\usepackage{courier}  % DO NOT CHANGE THIS
\usepackage[hyphens]{url}  % DO NOT CHANGE THIS
\usepackage{graphicx} % DO NOT CHANGE THIS
\usepackage{graphicx}  % DO NOT CHANGE THIS
\usepackage{multirow}
\usepackage{mathastext}

\title{Modeling Institutional Credit Risk with Financial News}
\author{Tam Tran-The\\ % All authors must be in the same font size and format. Use \Large and \textbf to achieve this result when breaking a line
MassMutual Data Science\\ %If you have multiple authors and multiple affiliations
470 Atlantic Ave\\
Boston, Massachusetts 02210\\
ttranthe32@massmutual.com % email address must be in roman text type, not monospace or sans serif
}
 
\begin{document}

\maketitle
%------------------------------------------------------------------------- 
\begin{abstract}
Credit risk management, the practice of mitigating losses by understanding the adequacy of a borrower's capital and loan loss reserves, has long been imperative to any financial institution's long-term sustainability and growth. MassMutual is no exception. The company is keen on effectively monitoring downgrade risk, or the risk associated with the event when credit rating of a company deteriorates. Current work in downgrade risk modeling depends on multiple variations of quantitative measures provided by third-party rating agencies and risk management consultancy companies. As these structured numerical data become increasingly commoditized among institutional investors, there has been a wide push into using alternative sources of data, such as financial news, earnings call transcripts, or social media content, to possibly gain a competitive edge in the industry. The volume of qualitative information or unstructured text data has exploded in the past decades and is now available for due diligence to supplement quantitative measures of credit risk. This paper proposes a predictive downgrade model using solely news data represented by neural network embeddings. The model standalone achieves an Area Under the Receiver Operating Characteristic Curve (AUC) of more than 80\%. The output probability from this news model, as an additional feature, improves the performance of our benchmark model using only quantitative measures by more than 5\% in terms of both AUC and recall rate. A qualitative evaluation also indicates that news articles related to our predicted downgrade events are specially relevant and high-quality in our business context. 
\end{abstract}

%------------------------------------------------------------------------- 
\section{Introduction}
Credit risk refers to the possibility of loss resulting from a borrower's or a bond issuer's failure to repay a loan or meet contractual obligations. One subcategory of credit risk is downgrade risk, which occurs when third-party rating agencies, such as Moody's and S\&P, lower their ratings on a bond or a company. For example, a change by S\&P from a B to a CCC rating is considered a downgrade event. Such rating information is used extensively by regulatory organization, specifically the National Association of Insurance Commissioners (NAIC), to ensure financial solvency of insurance companies. More precisely, NAIC requires a company with a higher exposure to risk to hold a higher amount of capital in reserve. Accurately monitoring rating classes and potential deterioration events is therefore critical to MassMutual. Receiving enough notice before any possible impending downgrade event would help the company to manage its investment portfolio more efficiently by preparing necessary capital and/or switching holding positions. 

In this paper, we design 3 natural language processing (NLP) frameworks to recognize credit relevant patterns (i.e. future 1-year downgrade events) in news articles about more than 2.2K companies of interest. The study demonstrates that with an appropriate methodology, we can achieve an AUC of more than 80\% using solely news data to predict downgrade events and yield a more than 5\% performance gain when this adverse credit signal in media coverage is added to the quantitative downgrade risk model. 

The structure of this paper is as follows: Section \ref{background} provides background on credit risk modeling and the NLP techniques used to convert text into meaningful numerical data; Section \ref{main development} describes the dataset, benchmark model and main model development; Section \ref{results} presents performance results and stability test details of the final model; and Section \ref{conclusions} discusses limitations of this work and future directions.  

%------------------------------------------------------------------------- 
\section{Background}\label{background}
%------------------------------------------------------------------------- 
\subsection{Credit Risk Modeling}
The motivation to develop credit risk models stems from the need to construct quantitative estimates of the amount of economic capital needed to support a financial institution's risk-taking activities. Specifically, minimum capital in reserve is often set in proportion to the risk exposure of a company's portfolio. Typical credit risk models take as input the conditions of the general economy and those of the specific firms in question, and generate as output a credit quality measure. 

%------------------------------------------------------------------------- 
\subsection{Natural Language Processing (NLP) Techniques: Topic Modeling, Sentiment Lexicons and Neural Net Embeddings}
When it comes to using text in a machine learning model, one of the main challenges is how to represent texts as numerical inputs so that we can feed them into the model. This project aims to experiment with different methodologies to represent news articles and evaluate the algorithms' performances based on the downgrade prediction task. Following are some NLP techniques we use in the study: 

\textbf{Latent Dirichlet Allocation (LDA)}. As it is believed that every article we have is composed of major themes, we use topic modeling to extract the hidden thematic structure in the text. Topic modeling is a type of statistical model used to discover the abstract topics that occur in a collection of documents. Latent Dirichlet allocation (LDA), a generative probabilistic model, is a common algorithm of topic model. The basic idea is documents are represented as random mixtures over latent topics, where each topic is characterized by a distribution over words. Mathematically, this can be formulated as:
\[ P(\theta, \textbf{z}, \textbf{w} | \alpha, \beta) = P(\theta | \alpha) \prod_{n=1}^{N} P(z_n|\theta)P(w_n|z_n, \beta) \]
Given $\alpha$, a parameter vector on the per-document topic distributions and $\beta$, a parameter vector on the per-topic word distributions, we are finding the joint distribution of a topic mixture $\theta$, a set of N topics \textbf{z}, and a set of N words \textbf{w} \cite{Blei03latentdirichlet}. In the context of our study, each news article is represented as a set of topic probabilities and the number of topics is a tunable parameter. To evaluate an LDA model, we use coherence measure, which gives an estimate of how well each topic can be represented as a composition of parts that can be combined \cite{Roder:2015:EST:2684822.2685324}. 

\textbf{Sentiment Lexicons}. There is a growing body of sentiment lexicons, or affective word lists, to examine the tone and sentiment of textual data. Some lexicons we experiment with are: 
\begin{itemize}
    \item \textbf{Loughran and McDonald}, designed to particularly reflect tone in financial texts \cite{RePEc:bla:jfinan:v:66:y:2011:i:1:p:35-65}
    \item \textbf{VADER}, specifically attuned to sentiment in microblog-like contexts \cite{HuttoG14}
    \item \textbf{AFINN}, also constructed based on micro-blog posts including Internet slang words \cite{DBLP:journals/corr/abs-1103-2903} 
    \item \textbf{SentiWordNet} and \textbf{OPINION}, for general sentiment analysis
\end{itemize}

\textbf{Neural Net Embeddings}. Based on the underlying idea that ``a word is characterized by the company it keeps,'' each word is represented by an embedding, or a vector of continuous numbers so that words of similar semantic (relative to the task) are closer to one another in the vector space. By using neural network on a supervised/unsupervised task, the resulting weights/parameters that have been adjusted to minimize loss on the task are the embeddings. 

\begin{itemize}
    \item \textbf{Doc2Vec Document Embeddings}. Doc2Vec is an unsupervised framework that learns fixed-length feature representations from variable-length pieces of texts. In the framework, a vector representation is trained by stochastic gradient descent and backpropagation to be useful for predicting the next word in the sentence. The Doc2Vec model we use in this study is a distributed memory one, where each paragraph vector acts as a memory that remembers what is missing from the current context \cite{Le:2014:DRS:3044805.3045025}. After being trained, the paragraph vectors can be used as features for the paragraph and be fed directly into conventional machine learning algorithms such as logistic regression. 
    \item \textbf{fastText Word Embeddings}. Inspired by the same hypothesis as Doc2Vec, fastText word representations are trained to predict words that appear in their contexts. What distinguishes fastText model is it takes into account morphology. More precisely, the model represents each word by a sum of its character n-grams, which allows us to compute representations for words that did not appear in the training data and proves to be helpful while working with morphologically rich language \cite{DBLP:journals/corr/BojanowskiGJM16}.
\end{itemize}

%------------------------------------------------------------------------- 
\section{Modeling Downgrade Risk}\label{main development}
%------------------------------------------------------------------------ 
\begin{figure*}[t!]
    \center{\includegraphics[width=0.8\textwidth]
    {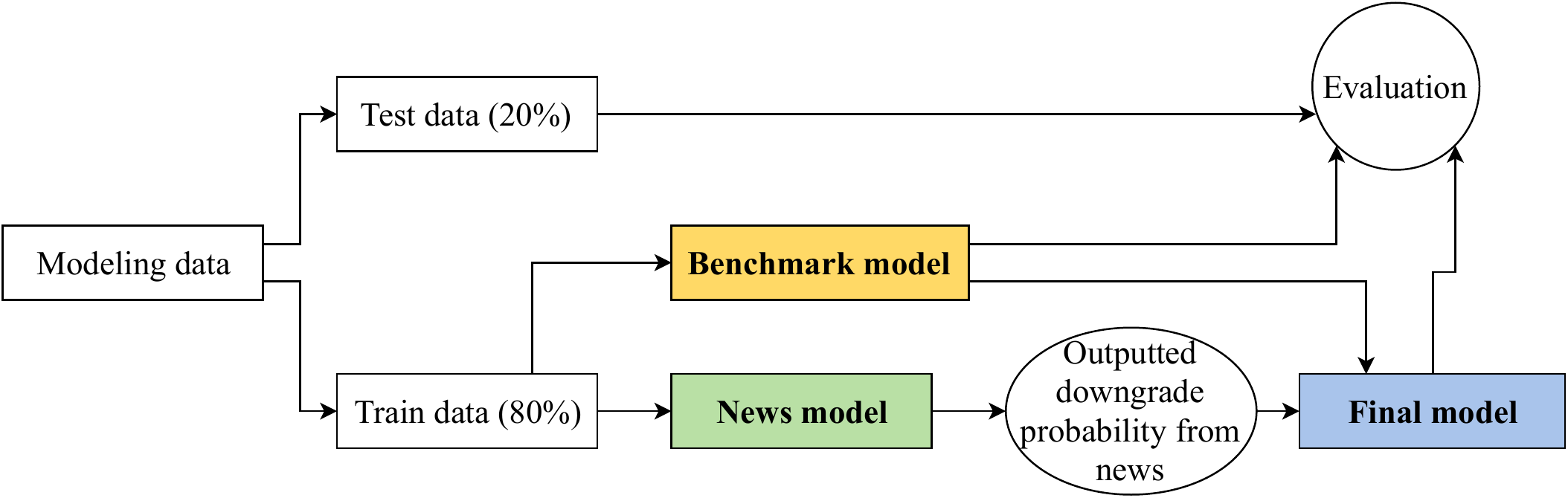}}
    \caption{\label{fig:modeling-pipeline} Overview of modeling pipeline. For details of benchmark model, see Section \ref{benchmark}. For details of news model, see Section \ref{maindev}. Note that features for the final model include the outputted probability from news model and all quantitative features from benchmark model.}
\end{figure*}

\subsection{Data}
 
\textbf{News}. For each company at a point in time (daily), we have all news articles corresponding to the record and a label indicating whether the company is going to downgrade within future 1-year period. 

\begin{itemize}
    \item \textbf{Data source}. We have access to news data source from Thomson Reuters via CreditEdge's API provided by Moody's. This source brings us daily article pieces at company level. Reuters articles in the dataset could take various forms: news stories about company's activities (e.g. merger \& acquisition, board assignment, etc.), market snapshots (e.g. individual stock movements, etc.), or quarterly earnings summaries. There are approximately 21K observations identified at company and date level, covering more than 2.2K Moody's unique permanent company identifiers (PID) and spanning from mid November 2017 to early October 2019 (although 99\% of these articles are in 2019).
    \item \textbf{Data preprocessing}. Since we are working in a very narrow domain (financial news of a company of interest) and the data is particularly noisy, we take multiple cleaning and preprocessing steps. First, we remove small machine-generated articles, such as NYSE/NASDAQ/AMEX order imbalance, articles that only include video links or articles whose accuracy hasn't been verified by Reuters. Second, we remove unnecessary headers, footers, HTML metadata, and lowercase all texts. Third, we manually read through a handful of articles and flag which format tends to mention multiple companies in the text and which tends to talk about a single company. Last but not least, for pieces that cover information of multiple companies, we use a combination of named entity recognition and fuzzy matching techniques, as well as rule-based exact matching to extract only sentences about the company of relevance. A few more preprocessing layers, such as tokenization, stemming and stop words removal, are added depending on which modeling approach we take. By performing a thorough text preprocessing, we hope to increase the signal to noise ratio which suggests news is more likely to have a material impact on our downgrade prediction. To ensure there is only one row of data per company per day, for companies that have multiple articles on a single day, we concatenate these articles together. 
\end{itemize} 

\textbf{Ratings}. Credit ratings for each company are available from both Moody's and S\&P. We take the worst combination between the two rating systems for a company at a point in time (daily) to derive current rating and the worst rating within the next future 1 year, from which we determine whether that company has an impending downgrade event or not.

\begin{table}[h]
\caption{Downgrade events by year}\smallskip
\centering
\begin{tabular}{l|l|l}
downgrade & year & count \\ \hline
\multirow{3}{*}{0} & 2017          & 42             \\ 
                   & 2018          & 152            \\ 
                   & 2019          & 20,519         \\ \hline
1                  & 2019          & 324            \\ 
\end{tabular}
\label{tab:dg-count-yearly}
\end{table}

For our benchmark model, all features are variations of quantitative credit measures provided by third-party rating agencies and risk management consultancy companies - this is further discussed in Section \ref{benchmark}. 

%------------------------------------------------------------------------- 
\subsection{Benchmark Model} \label{benchmark}
The benchmark downgrade model of this study is a logistic regression. For each training data point, we have a vector of features $X$ and an observed class $Y$, where $X$ provides quantitative metrics about a company on a daily basis and $Y=1$ indicates that the company is downgrading within 1 year from the date of the data observation. This setup ensures that both our benchmark and news models have the same level of data frequency and can be compared and combined later. Assuming that $P(Y=1|X=x) = p(x;\theta)$ for some function $p$ parameterized by $\theta$, what we want to model is: 

\[ P(Y=1|X=x, \theta) = \frac{1}{1 + e^{-\theta x}} \]

The model is trained on 9 features that are variations in terms of either term structures (i.e. 1-year, 5-year) or transformation (i.e. lag, diff) of following quantitative measures purchased from risk management providers: 
\begin{itemize}
    \item The probability that a firm will default over 1 year based on company-specific attributes, industry-related measures and relevant macro-economic factors 
    \item The probability that a firm will downgrade within 1 year based on market implied ratings and rating outlooks
    \item Historical credit rating for any firm at a point in time
\end{itemize}

Since our data is highly imbalanced, where the occurrence of downgrade events accounts for 1.5\% of the dataset, we employ SMOTE algorithm to replicate observations from the minority class. Overall, this benchmark model achieves an AUC of 82.7\% and a recall rate of 69.5\%.

%------------------------------------------------------------------------- 
\subsection{Model Development} \label{maindev}

Figure \ref{fig:modeling-pipeline} is an overview of our modeling pipeline. Throughout the training process, we deliberately employ logistic regression due to the algorithm's great interpretability which is a focus of our business users who work in a highly regulated industry. This section details 3 different approaches we take to translate unstructured texts into meaningful numerical data that can be fed into a logistic regression model. We evaluate these NLP methodologies on our downstream task at hand, which is downgrade prediction, to select the best news model that can be incorporated into and improve the existing benchmark model. 

\begin{figure}[h]
    \centering{\includegraphics[width=0.9\columnwidth]
    {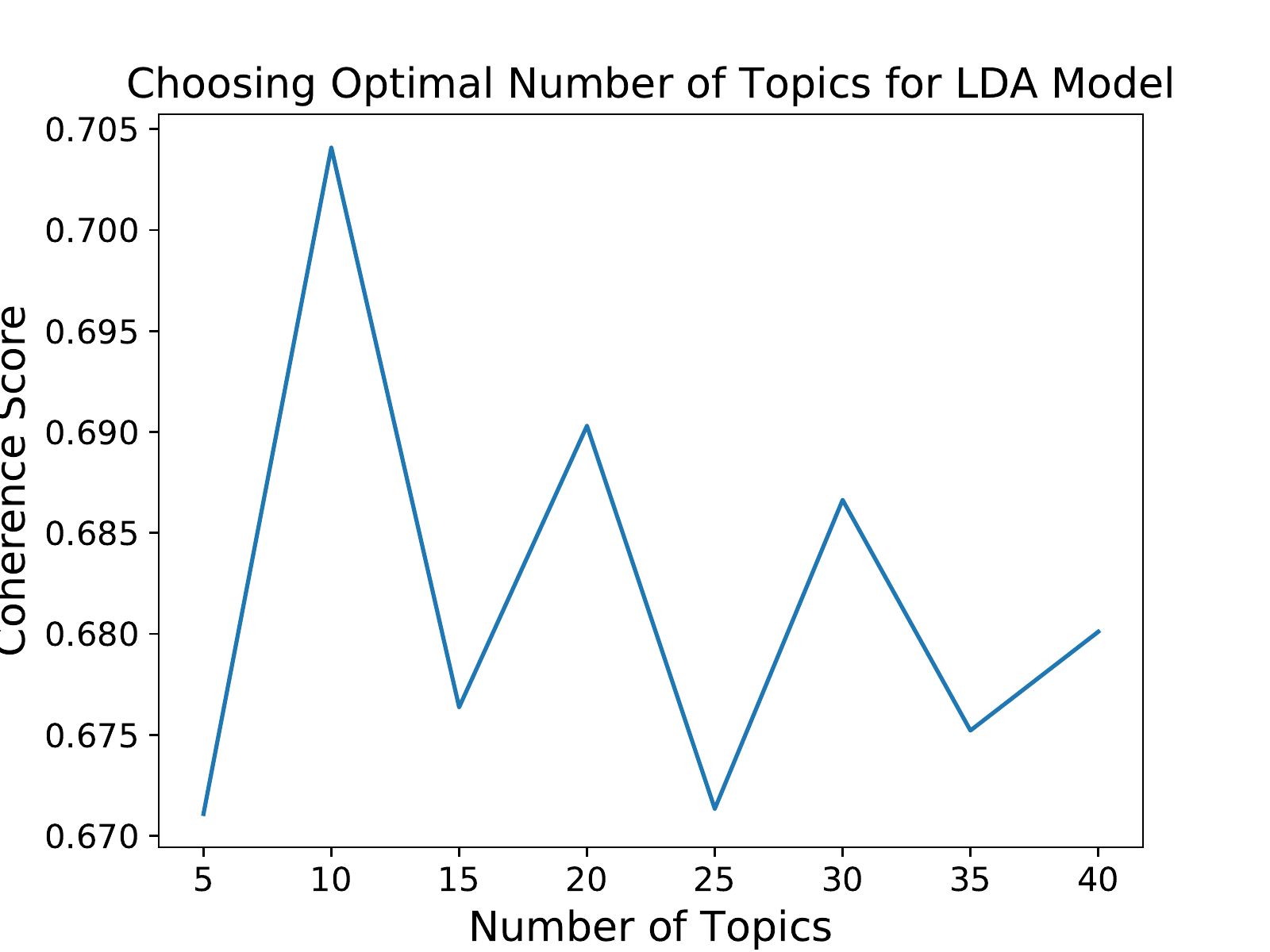}}
    \caption{\label{fig:coherence} Coherence values of LDA model with different topic numbers. The optimal number is 10 topics with coherence value of 0.704.}
\end{figure}

\textbf{Approach 1: Lexicon-based Sentiment and Topic Scores}. We first experiment with one of the simplest sentiment analysis approaches, which is to compare words of an article against a labeled word list, where each word has been scored for valence. Based on 5 different lexicons (Loughran \& McDonald, VADER, AFINN, SentiWordNet, and OPINION), we compute 5 sentiment scores for each news article. Specifically, we check whether a word in the text exists in each positive or negative word list, then count the frequency of that word. For positive words, we do not count those that have a negation in one of three places preceding it. The sentiment score is calculated as $\frac{\# positives - \# negatives}{\# positives + \# negatives}$. Additionally, we train an LDA model with 10 different topics where each topic is a combination of keywords and each keyword contributes a certain weight to the topic. We decide on 10 as the optimal number of topics by running multiple LDA models with number of topics ranging from 5 to 40 and picking the one that gives the highest coherence value, as shown in Figure \ref{fig:coherence}. Two features that indicate the number of articles for each company at a point in time and whether an article contains any variation of the word ``downgrade'' are also created. In the end, our news model is trained on 17 features, including 5 sentiment scores, 10 topic probabilities, and 2 count indicators. This news model standalone has an unimpressive AUC of 59.8\%. Although sentiment and topic scores are helpful in giving us a basic analysis of our news data at hand, these features are far from sufficient for accurate downgrade prediction. 

\textbf{Approach 2: Doc2Vec Embeddings}. 
In this approach, we train a distributed-memory Doc2Vec model based on the vocabulary of our training dataset over 20 epochs and generate a document embedding for each news article. We choose to use Doc2Vec paragraph vectors since they are learned from unlabeled data and theoretically can work well for tasks that have small amount of labeled data \cite{Le:2014:DRS:3044805.3045025}, which is our case in this study. These embeddings are taken as input into a logistic regression model, which results in an AUC of 71.6\%. 

\textbf{Approach 3: fastText Embeddings}. 
Taking advantage of transfer learning, we use 1M word vectors pre-trained by a team of Facebook researchers on Wikipedia 2017, UMBC webbase corpus and statmt.org news dataset \cite{mikolov2018advances}. Since the fastText model was trained on massive data sources, their representations perform very well at transferring to other NLP problems and improve the generalization of models learned on limited amount of data. Unlike Doc2Vec which produces representations at document level, fastText generates a 300-dimensional vector for each word. To create an embedding for the whole document, we take the average of embeddings for words contained in it. With these final document embeddings fed into a logistic regression, our news model achieves an AUC of 80.9\%, which is the best performance for this standalone model so far. 

\begin{table}[h]
\caption{Performance of standalone news model} \smallskip
\centering
\begin{tabular}{l|l}
approach                   & AUC (\%) \\ \hline
sentiment scores and LDA topic scores & 59.8                          \\ 
Doc2Vec embeddings                    & 71.6                          \\ 
fastText embeddings (averaging out)   & \textbf{80.9}                          \\ 
\end{tabular}
\label{tab:news-auc}
\end{table}

Based on the performance results, we decide to employ fastText embeddings as the main feature of our news model. The downgrade probability outputted from this is then combined with other quantitative measures in the benchmark, coupled with SMOTE oversampling technique, to build a final augmented model. Results of this final model are further explored in Section \ref{results}.
%------------------------------------------------------------------------- 
\section{Final Model Results and Validation}\label{results}

\subsection{Quantitative Evaluation}\label{quanteval}
The following results are based on the evaluation on a 20\% hold-out test set that are untouched, unseen throughout the training process. Both the benchmark and final models are tested on this same set to ensure fair comparison. 

\begin{figure}[t]
    \center{\includegraphics[width=0.9\columnwidth]
    {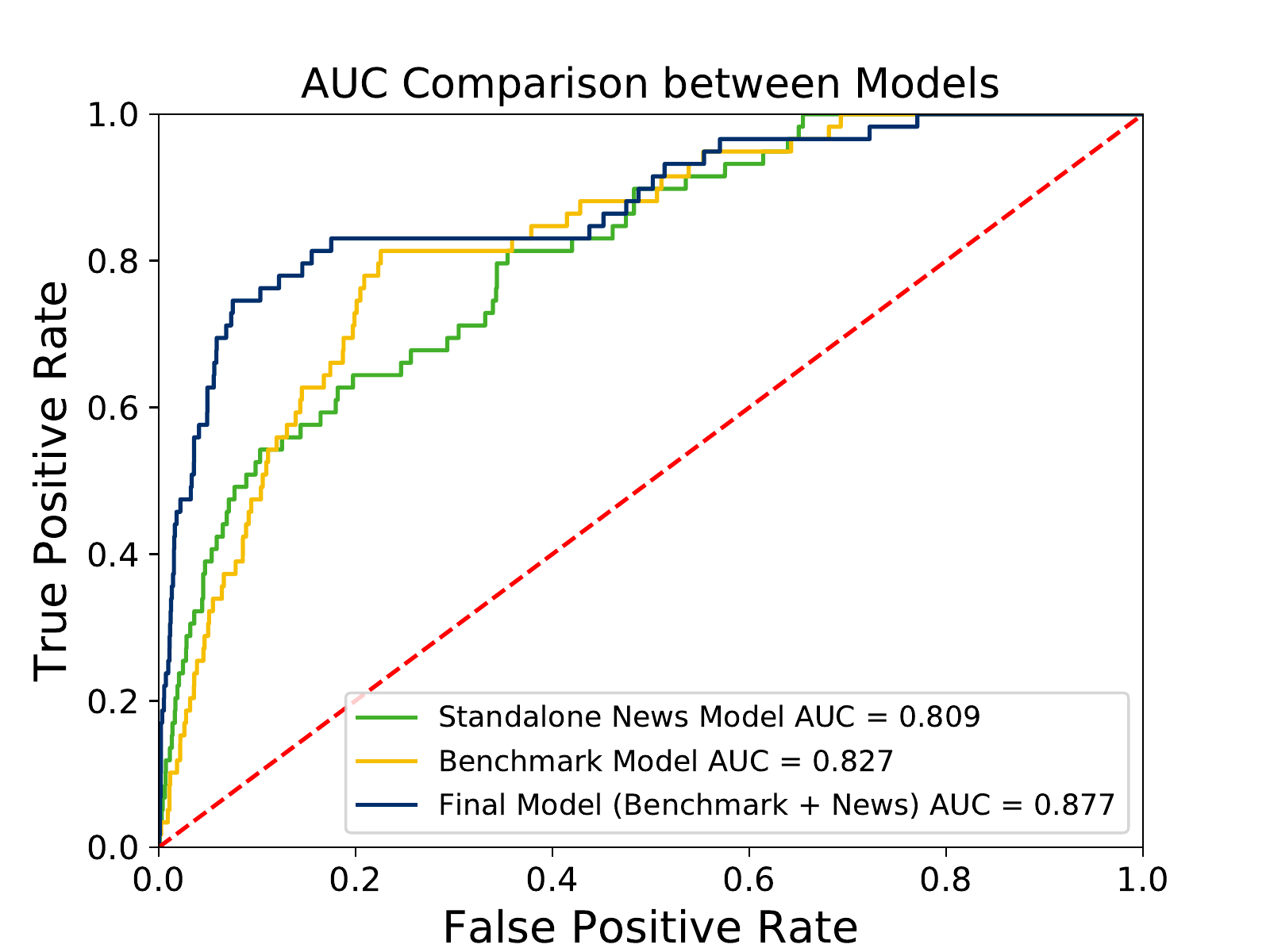}}
    \caption{\label{fig:auc} AUC comparison between benchmark, standalone news, and final models. Final model achieves the highest value, 87.7\%, resulting in a 5\% gain compared with benchmark model.}
\end{figure}

\textbf{AUC}. One of the main metrics we use to compare performances of different models is AUC, which tells how good a model is at distinguishing classes (the higher the number, the better). Figure \ref{fig:auc} indicates that a downgrade model using news alone can achieve a test AUC of 80.9\% and there is a 5\% gain when adding this news probability on top of the existing model (from 82.7\% for benchmark to 87.7\% for final model).

\textbf{Recall}. Since the cost of a false negative in our business context (an impending downgrade event that the model is not able to capture) is very steep, we put an emphasis on optimizing the model's recall rate. Although a standalone news model produces a humble recall rate of 54.2\%, incorporating news into benchmark model improves recall rate from 69.5\% to 74.6\% on the test set, resulting in a 5.1\% gain in recall. 

\begin{figure}[t]
    \center{\includegraphics[width=0.9\columnwidth]
    {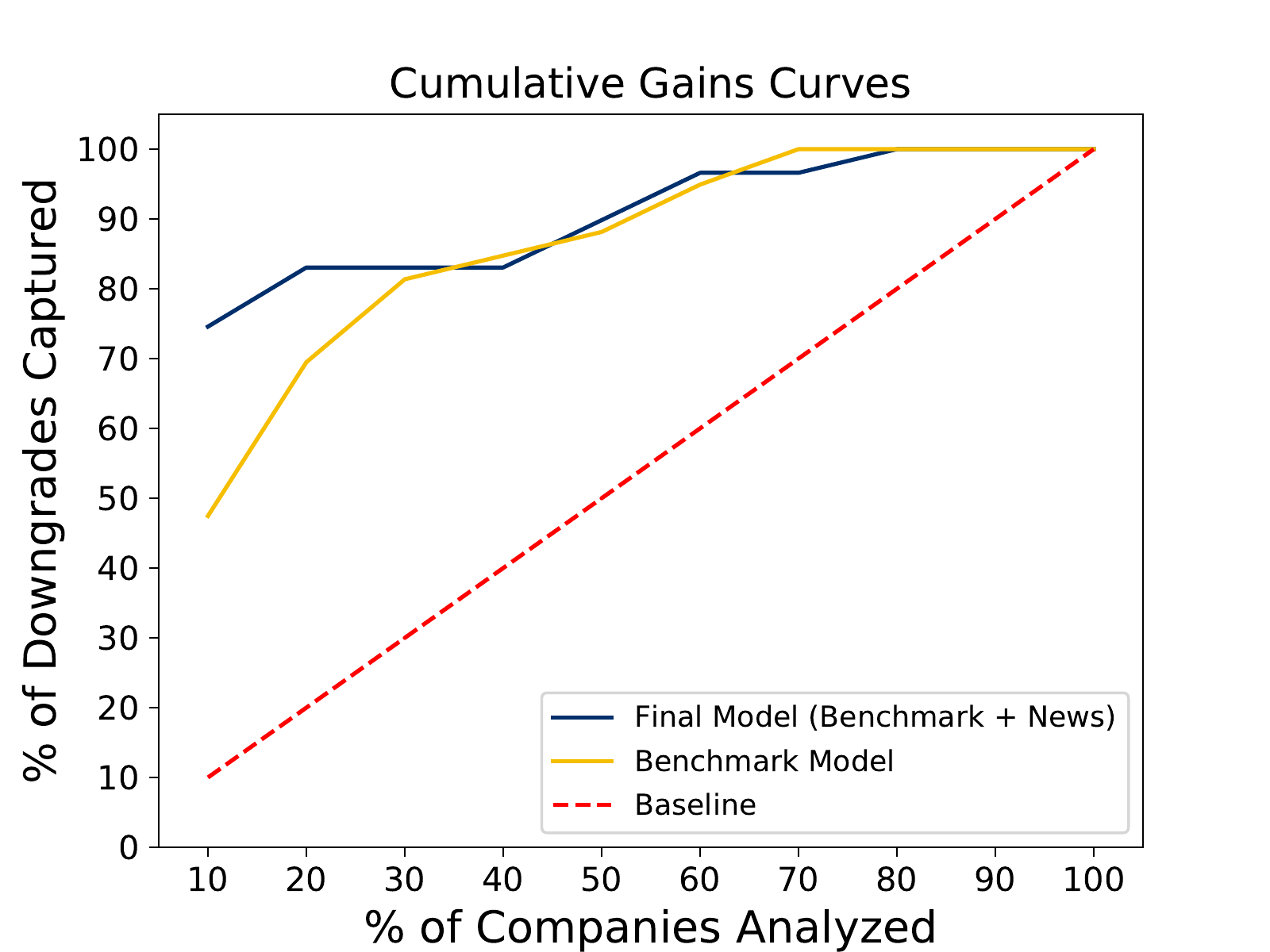}}
    \caption{\label{fig:lift} Cumulative gains chart. Given limited resources, the final model is able to provide a more optimal list of companies to analyze.}
\end{figure}

\textbf{Cumulative Gains}.
Given limited human and time resources that can be dedicated to the task of analyzing companies in the list of anticipated downgrades, we would like to be able to capture as many downgrades as possible using as few test cases as possible. Figure \ref{fig:lift} indicates that with news information added, analyzing the top 10\% companies with the highest predicted downgrade probabilities can achieve a recall rate of 74.6\%, which is a 27.1\% improvement compared with the benchmark model. Similarly, analyzing the top 20\% companies using the final model results in a recall rate of 83.1\%, an increase of 13.6\% compared with benchmark. 

Overall, our model with the additional downgrade probability of news not only has higher accuracy rate (in terms of AUC and recall) but is also more efficient by providing a more optimal list of companies to analyze given restrained resources.  

\subsection{Qualitative Evaluation}

To enhance the pragmatic sense of our model, we carefully examine all news articles associated with the true positives (actual downgrade events that the model is \textit{able} to capture) and false negatives (actual downgrade events that the model is \textit{unable} to capture) in our test set. Investigation shows that the model can pick up high-quality and very relevant article pieces to infer a downgrade event of a company.

\textbf{True Positive Samples}.
There are 44 true positives corresponding with 16 unique companies in our test data. Following is a representative list of articles appearing in our true positive cases. Each example is shown at company level, contains only keywords due to space limit, and gives a good summary of the entire article piece.

\begin{itemize}
    \item Company A: close factories; cut down 12,000 jobs; has been among the hardest to be hit by the trade war so far
    \item Company B: lose clients and get sued following misconduct revelations
    \item Company C: challenged by [an investor] for stretching itself financially to buy rival oil driller 
    \item Company D: struggle with a host of issues; cut its dividend and report a wider-than-expected loss in its main engineering and construction unit
    \item Company E: hire restructuring firms and may choose to seek bankruptcy protection
    \item Company F: plan to wind down its dress-barn retail operations, resulting in the closure of about 650 stores
    \item Company G: anticipate having discussions on constructive basis relating to its underperformance
    \item Company H: make sophisticated missiles that use rare earth metals in their guidance systems, and sensors
    \item Company I: face lawsuit over art fraud; had been the willing auction house that knowingly and intentionally made the fraud possible
\end{itemize}

Although none of the articles explicitly mention that a company is being downgraded, the majority of them provide valuable insights into a company's financial health or how the company is perceived in the market. These signals could potentially be considered as leading indicators toward a downgrade. The model seems to be able to pick up subtle and indirect signs of an impending downgrade event as well. For example, in the case of Company H, the article refers to the company's heavy dependence on rare earth metals. This does not appear to relate to downgrade at first glance; however, more examination suggests that rare earth metal has become a political spotlight recently due to the trade tension between the U.S, who considers this mineral critical to the country's economic and national security, and China, who is the largest producer and manufacturer of this element in the world. Company H indeed was downgraded on 07-31-2019.

\textbf{False Negative Samples}.
The test set includes 15 false negatives associated with 6 distinct companies. After review, we see these cases fall under either of two following categories:
\begin{itemize}
    \item The news articles follow unpopular format structures compared with others in the dataset, which leads to their noisy quality even after preprocessing.
    \item The news articles appear in this false negative list are about companies that also exist in the true positive list. However, these pieces have a positive/neutral tone, were published shortly earlier, and might revolve around a different activity mentioned in the articles of the true positives. For example: 
    \begin{itemize}
        \item Company I's shareholders approve proposed acquisition; clients are pleased with the company's help to file their first lawsuit against the government (This piece was published on 06-21-2019. The one in true positive list was published on 06-25-2019. Company I's downgrade event happened on 09-17-2019).
        \item Company C could divest most or all of [an investment vehicle] after buying out [another company]. (This piece was published on 06-14-2019. The one in true positive list was published on 06-27-2019. Company C's downgrade event happened on 08-01-2019).
    \end{itemize}
    Our takeaways here are that bad news can ``outweigh'' good/neutral news when it comes to contributing to the prediction of downgrade and companies' ratings could take downturn very shortly after emergence of some negative sentiments about them in the market.  
\end{itemize}

\subsection{Model Robustness}
\begin{figure}[t]
    \center{\includegraphics[width=0.9\columnwidth]
    {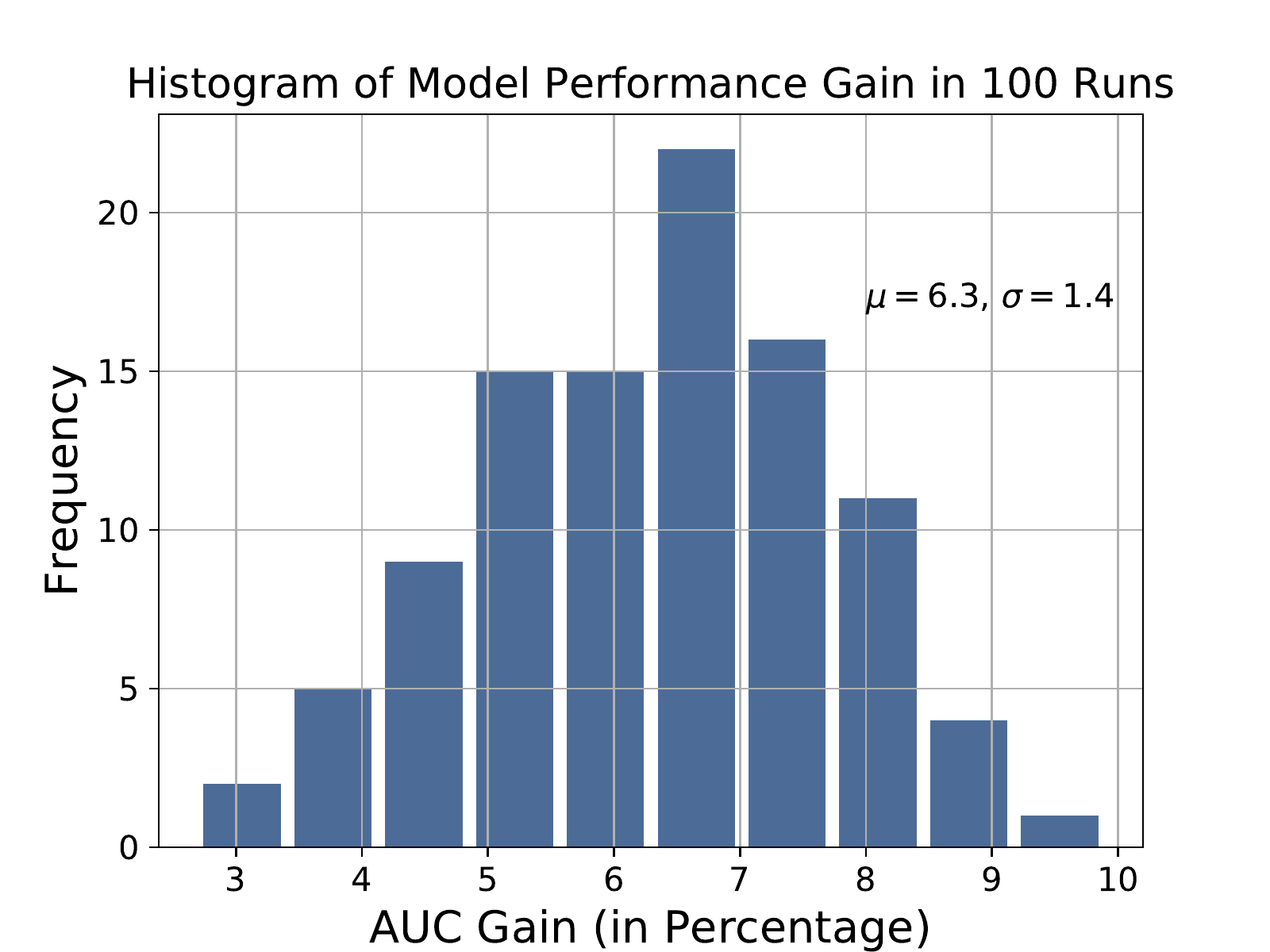}}
    \caption{\label{fig:100expshist} Histogram of the performance gain in 100 experiment runs. The gain is positive 100 times, with mean value 6.3\% and standard error 1.6\%.}
\end{figure}

To evaluate the robustness of our final model's performance gain, we run 100 experiments using 100 different random seeds that are involved in the training/splitting, cross-validation and SMOTE oversampling processes. The performance gain in AUC from adding news to the benchmark model is positive 100 times out of 100 experiments. The mean value of these gains is 6.3\% with a standard error of 1.6\%, as shown in Figure \ref{fig:100expshist}.

This suggests that our positive performance gain is reliable and the 5\% increase in AUC mentioned in Section \ref{quanteval} is still on the lower side. As we have access to a larger set of data, the standard error will theoretically be pushed closer to 0. 
%------------------------------------------------------------------------- 
\section{Conclusions and Future Work}\label{conclusions}
In the hope of obtaining unique insights into companies' financial health that are not available in traditional quantitative credit measures, we train a downgrade risk model on solely news information. We demonstrate that news coverage, if represented appropriately in the data, can help detect adverse credit signal and considerably improve the performance of existing model that is trained on conventional credit measures. 

There are multiple avenues for research opened from this work. Throughout our modeling development, we notice how we preprocess the text data could introduce a material change in our final model performance. Extracting only relevant information about the company of interest plays a key role in increasing the gain. Thus, we would like to further explore more advanced techniques regarding information extraction, such as coreference resolution to find sentences that do not exactly include a company's name but still refer to the same entity, or sentence segmentation to pick out sentences when the boundary is ambiguous. In addition, because the news model is built solely based on online articles, it is inherently subject to media bias. Fact-checking news or tackling different forms of bias within mass media is outside the scope of this study but offers an exciting opportunity for future research. 

%------------------------------------------------------------------------- 
\section{Acknowledgement}
The author is grateful for helpful discussions about model development made by Zizhen Wu and Jasmine Geng, for knowledge sharing about the raw datasets made by Yi Li and Yi Wang, and for paper review by Nailong Zhang, Jasmine Geng, Adam Fox and Sears Merritt. 
%------------------------------------------------------------------------- 
\bibliographystyle{aaai}
\bibliography{bibliography}
\end{document}